\documentclass[reprint,aps,prx,superscriptaddress,notitlepage,nopacs,amsmath,amstex,amssymb,citeautoscript,longbibliography,floatfix,letter,10pt]{revtex4-2}
\usepackage{graphicx} 
\usepackage{amsmath}
\usepackage{braket}
\usepackage{tabularx}
\usepackage{multirow}
\usepackage[dvipsnames]{xcolor}
\usepackage[bookmarks=true,colorlinks,linkcolor=OrangeRed,urlcolor=NavyBlue,citecolor=RoyalBlue]{hyperref}
\usepackage{float}
\usepackage{comment}

\begin{document}

\title{
Interplay of Confinement and Localization in a Programmable Rydberg Atom Chain
}

\author{Andrea B. Rava}
    \altaffiliation{a.rava@fz-juelich.de}
	\affiliation{Jülich Supercomputing Centre, Forschungszentrum Jülich, D-52425 Jülich, Germany}
    \affiliation{RWTH Aachen University, 52056 Aachen, Germany}
  \author{J. A. Monta\~nez-Barrera}
    \altaffiliation{j.montanez-barrera@fz-juelich.de}
	\affiliation{Jülich Supercomputing Centre, Forschungszentrum Jülich, D-52425 Jülich, Germany}
    
\author{Kristel Michielsen}
	\affiliation{Jülich Supercomputing Centre, Forschungszentrum Jülich, D-52425 Jülich, Germany}
	\affiliation{Department of Computer Science, University of Cologne, 50931 Cologne, Germany }
\author{Jaka Vodeb}\email{jaka.vodeb@ijs.si}
\affiliation{Jülich Supercomputing Centre, Institute for Advanced Simulation, Forschungszentrum Jülich, 52425 Jülich, Germany}
\affiliation{Department of Complex Matter, Jožef Stefan Institute, Jamova 39, 1000 Ljubljana, Slovenia}
\affiliation{CENN Nanocenter, Jamova 39, 1000, Ljubljana, Slovenia}

\date{\today}

\begin{abstract}
Analog quantum simulators promise access to complex many-body dynamics, yet their performance is ultimately set by how device imperfections compete with intrinsic physical mechanisms.  
Here we present an end-to-end study of correlation spreading in a programmable Rydberg-atom chain realizing a longitudinal-field transverse-field Ising model, focusing on the joint impact of confinement and effective disorder.  
Experiments performed on QuEra’s \emph{Aquila} quantum processor are benchmarked against large-scale coherent emulations using the Jülich Quantum Annealing Simulator (JUQAS), enabling the controlled inclusion of realistic hardware imperfections.  
In the ideal coherent limit, a tunable longitudinal field induces confinement of domain-wall excitations into mesonic bound states, leading to a progressive truncation of the correlation light cone.  
When experimentally relevant inhomogeneities and fluctuations are included, correlations instead saturate at finite distance even in the nominally deconfined regime, revealing localization driven by emergent disorder.  
The close quantitative agreement between noisy emulations and experimental data allows us to attribute the observed saturation to specific hardware error channels and to identify the dominant contribution.  
Our results establish a practical framework for diagnosing and modeling error-induced localization in Rydberg quantum processors, while demonstrating that confinement remains a robust and programmable mechanism for engineering non-ergodic dynamics on near-term quantum hardware.
\end{abstract}

\maketitle

\section{Introduction}

The study of non-equilibrium quantum dynamics has revealed that ergodicity breaking, where a system fails to explore its entire Hilbert space, can arise through several distinct physical mechanisms. Among the most prominent are \emph{confinement}, which binds elementary excitations through interaction-induced potentials, and \emph{localization}, which suppresses transport due to disorder, inhomogeneity, or dynamical constraints. While both mechanisms restrict quasiparticle motion and preserve memory of the initial configuration, their microscopic origins differ: confinement is energetic, whereas localization is spatial. Understanding how these two forms of restricted dynamics interact is a complex open question at the intersection of condensed-matter physics, quantum information, and quantum simulation.

Rydberg atom arrays have recently emerged as powerful platforms for exploring strongly correlated spin dynamics in and out of equilibrium~\cite{browaeys2020many,bernien2017probing,bluvstein2021controlling,cheng2024emergent}. Their combination of long-range interactions, single-atom addressability, and flexible geometry enables the realization of programmable Ising-type Hamiltonians, where transverse and longitudinal fields can be tuned in real time. This tunability provides direct experimental access to confinement phenomena, originally predicted in the non-integrable Ising model by Kormos \emph{et al.}~\cite{kormos2017real}, where domain walls experience a linear potential and form bound mesonic states. Subsequent theoretical work demonstrated that such confinement leads to truncated correlation light-cones~\cite{rutkevich2008energy,liu2019confined,mazza2019suppression,scopa2022entanglement}, slow relaxation, and oscillatory entanglement dynamics—non-ergodic signatures that closely resemble localization even in the absence of disorder.

Parallel research has shown that localization itself can arise through a variety of mechanisms beyond traditional disorder. Many-body localization (MBL)~\cite{imbrie2016many,wei2018exploring,lee2017many} establishes that strong random fields can halt thermalization in interacting systems, while more recent studies have uncovered disorder-free or constraint-induced localization in clean frustrated magnets and kinetically constrained spin models~\cite{mcclarty2020disorder,ostmann2019localization,marcuzzi2017facilitation,dag2024emergent,viehmann2013quantum}. In Rydberg arrays, emergent inhomogeneity—caused by atomic motion, laser-frequency noise, or imperfect detuning control—acts as a natural source of such disorder, producing sub-ballistic correlation spreading and arrested dynamics even in nominally uniform systems.

Despite these advances, the relationship between confinement and localization remains largely unexplored experimentally. Confinement limits the separation of domain walls by an interaction-generated string tension, whereas localization pins excitations through static or dynamical inhomogeneity. Whether these effects compete or reinforce each other—yielding distinct dynamical regimes ranging from coherent meson oscillations to noise-induced freezing—has remained an open question.

In this work, we address this problem by combining quantum simulation on QuEra’s \emph{Aquila} Rydberg quantum processor \cite{wurtz2023aquilaqueras256qubitneutralatom} with large-scale coherent emulations using the Jülich Quantum Annealing Simulator (JUQAS). We realize a one-dimensional transverse-field Ising chain with tunable longitudinal bias, initiating quenches from an ordered product state and tracking the connected correlation dynamics in real time. The results suggest a continuous crossover from coherent confinement, characterized by oscillatory but spatially bounded correlations, to noise-induced localization, where correlation fronts saturate and the initial pattern persists. By introducing controlled noise into the JUQAS emulations, we identify spatial inhomogeneity of the longitudinal-field pattern, atomic motion, and laser-amplitude fluctuations as the dominant physical sources of localization on the device.

Our findings demonstrate that confinement and localization in Rydberg spin chains are not competing phenomena but rather cooperative mechanisms that jointly suppress quasiparticle mobility. Confinement imposes energetic constraints on domain-wall motion, while noise-induced inhomogeneity immobilizes the resulting bound states, leading to a unified picture of non-ergodic dynamics in programmable quantum simulators. This establishes Rydberg platforms as a versatile arena for exploring the continuum between coherent confinement and effective localization, bridging the physics of lattice gauge theories, many-body localization, and quantum information storage.

\section{Results}

\subsection{Mapping from the Rydberg Hamiltonian to the Ising model}

\begin{figure*}[t]
\centering
\includegraphics[width=\linewidth]{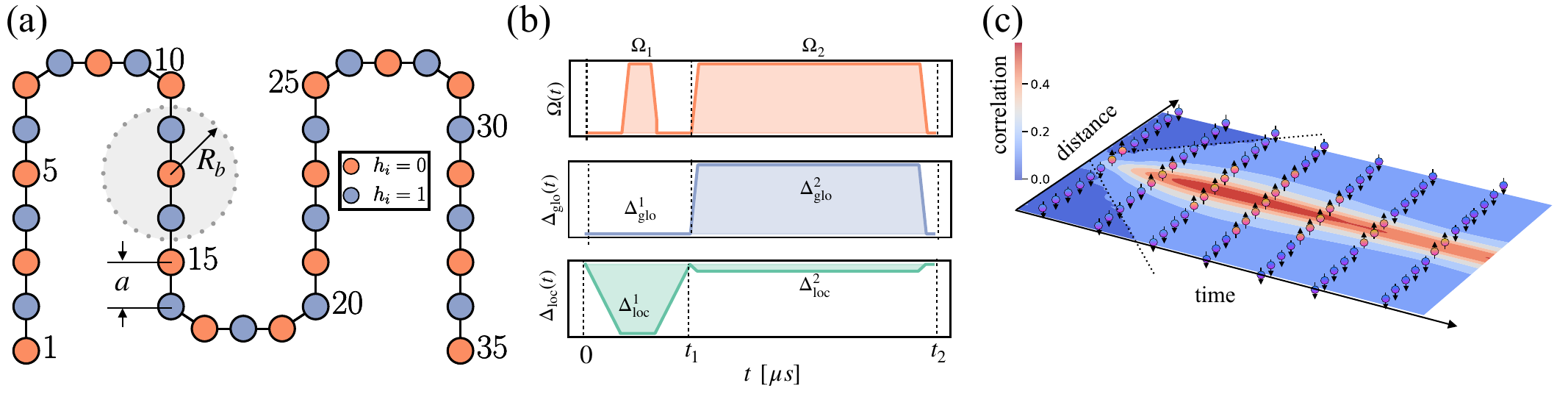}
\caption{
    \textbf{Experimental geometry, control sequence, and correlation spreading.}
    (a)~Schematic of the 35-atom Rydberg chain used in the experiment.  
    The atoms are arranged in a quasi-one-dimensional geometry with two $60^\circ$ turns to fit the optical-tweezer trapping region.  
    The lattice spacing $a$ and blockade radius $R_b$ determine the nearest-neighbour interaction scale.  
    During initialization, sites with $h_i = 0$ (orange) and $h_i = 1$ (blue) experience opposite local detunings $\pm \Delta_{\mathrm{loc}}$, preparing a Néel-ordered configuration.  
    (b)~Temporal pulse sequence applied in the experiment.  
    The global Rabi frequency $\Omega(t)$ (top) and detunings $\Delta_{\mathrm{glo}}(t)$ (middle) and $\Delta_{\mathrm{loc}}(t)$ (bottom) define two stages:  
    state preparation ($0 < t < t_1$) and quantum simulation ($t_1 < t < t_2$).  
    Dashed vertical lines mark the transition between stages, while shaded regions indicate finite quench ramps arising from hardware constraints.  
    (c)~Illustration of the measured correlation spreading dynamics.  
    The surface plot shows the time- and distance-resolved staggered connected correlations $(-1)^d C_{d(t)}$, with colour indicating correlation strength.  
    Superimposed spin configurations highlight the propagation and subsequent saturation of correlations during the quantum evolution.
    }
    \label{Fig:protocol}
\end{figure*}

The dynamics investigated in this work are governed by a Rydberg Hamiltonian that can be mapped onto an effective Ising model with tunable transverse and longitudinal fields.  
A schematic of the atomic register and control sequence is shown in Fig.~\ref{Fig:protocol}.  
The experiment employs $L = 35$ atoms arranged in a quasi-linear open chain with two $60^\circ$ turns to fit within the optical addressing region.  
During the initialization stage, alternating local detunings $\pm\Delta_\mathrm{loc}$ are applied to generate a Néel-ordered configuration, followed by a global quench of the drive amplitude $\Omega(t)$ and detunings $\Delta_\mathrm{glo}(t)$ and $\Delta_\mathrm{loc}(t)$ to initiate the dynamics.  
The temporal control sequence thus defines two stages: state preparation and quantum simulation.

The microscopic Hamiltonian describing the Rydberg array is
\begin{equation}\label{eq:HRyd}
\begin{split}
    \frac{H_\mathrm{Ry}}{\hbar} = -\frac{\Omega(t)}{2}\sum_{j}\left( e^{i\phi(t)}\ket{g_j}\bra{r_j} + e^{-i\phi(t)}\ket{r_j}\bra{g_j} \right) - \\
    - \sum_j\Delta_j(t)n_j + \sum_{j<k}U_{jk}n_jn_k,
\end{split}
\end{equation}
where $U_{ij}=C_6/|r_i-r_j|^6$ denotes the van der Waals interaction.  
Next, we substitute $n_i = (1+\sigma_i^z)/2$ and setting $\phi = 0$, yielding

\begin{equation}\label{eq:HRydsigmaz}
\begin{split}
    \frac{H_\mathrm{Ry}}{\hbar} = \sum_i\bigg [-\frac{\Omega(t)}{2}\sigma_i^x + \bigg(-\frac{\Delta_\mathrm {glo}(t)}{2} -\frac{\Delta_\mathrm {loc}(t)}{2}h_i\bigg)\sigma_i^z + \\
    + \sum_{j<i}\bigg(\frac{U_{ij}}{4}\sigma_i^z\sigma_j^z+\frac{U_{ij}}4{\sigma_i^z}+\frac{U_{ij}}4{\sigma_j^z}\bigg)\bigg].
\end{split}
\end{equation}

The final step in this mapping procedure is to explicitly incorporate the assumption that the atomic register satisfies  $U_{i,i+1} \gg U_{i,i+d}$ for $d>1$. Under these conditions, and considering a closed ring chain configuration, the Hamiltonian can be rewritten as

\begin{equation}\label{eq:HRydNN}
\begin{split}
    \frac{H_{\mathrm{Ry}}}{U_{i,i+1}/4} \approx \sum_i \bigg[-\dfrac{2\Omega(t)}{U_{i,i+1}}\sigma^x_i + \\
    +\bigg(-\dfrac{2\Delta_{\mathrm{glo}}(t) }{U_{i,i+1}} - \dfrac{2h_i\Delta_\mathrm{loc}(t) }{U_{i,i+1}} +2\bigg)\sigma^z_i + \sigma^z_i\sigma^z_{i+1}\bigg].
\end{split}
\end{equation}
In our analysis, we focus on the bulk of the Rydberg chain instead of the edge, thereby justifying the use of a closed chain Hamiltonian.

Because the Rydberg interaction is antiferromagnetic, a sublattice rotation $\sigma_i^z\!\rightarrow\!(-1)^{i+1}\sigma_i^z$ maps this Hamiltonian to the standard ferromagnetic Ising form
\begin{equation}
    \hat{H}_{\mathrm{Ising}} = -J\!\sum_i \!\left(h^x\sigma_i^x + h^z_i\sigma_i^z + \sigma_i^z\sigma_{i+1}^z\right),
    \label{eq:HIsiFM1}
\end{equation}
with $J=U_{i,i+1}/4$, $h^x = 2\Omega/U_{i,i+1}$, and $h^z_i = [2(\Delta_\mathrm{glo} + h_i\Delta_\mathrm{loc})/U_{i,i+1} - 2](-1)^{i+1}$.  
This mapping provides direct control over both transverse and longitudinal fields through the experimentally tunable drive and detuning parameters, enabling continuous interpolation between the deconfined ($h^z=0$) and confined ($h^z\neq0$) regimes of the Ising model. Most notably, we are able to make the longitudinal field homogeneous in the ferromagnetic interpretation of the Rydberg Hamiltonian, a critical requirement for realizing confinement dynamics in a ferromagnetic quantum Ising chain and we will refer to its magnitude as $h^z$.

\subsection{Confinement and noise-induced suppression of correlation spreading}

\begin{figure*}[t]
\centering
\includegraphics[width=\textwidth]{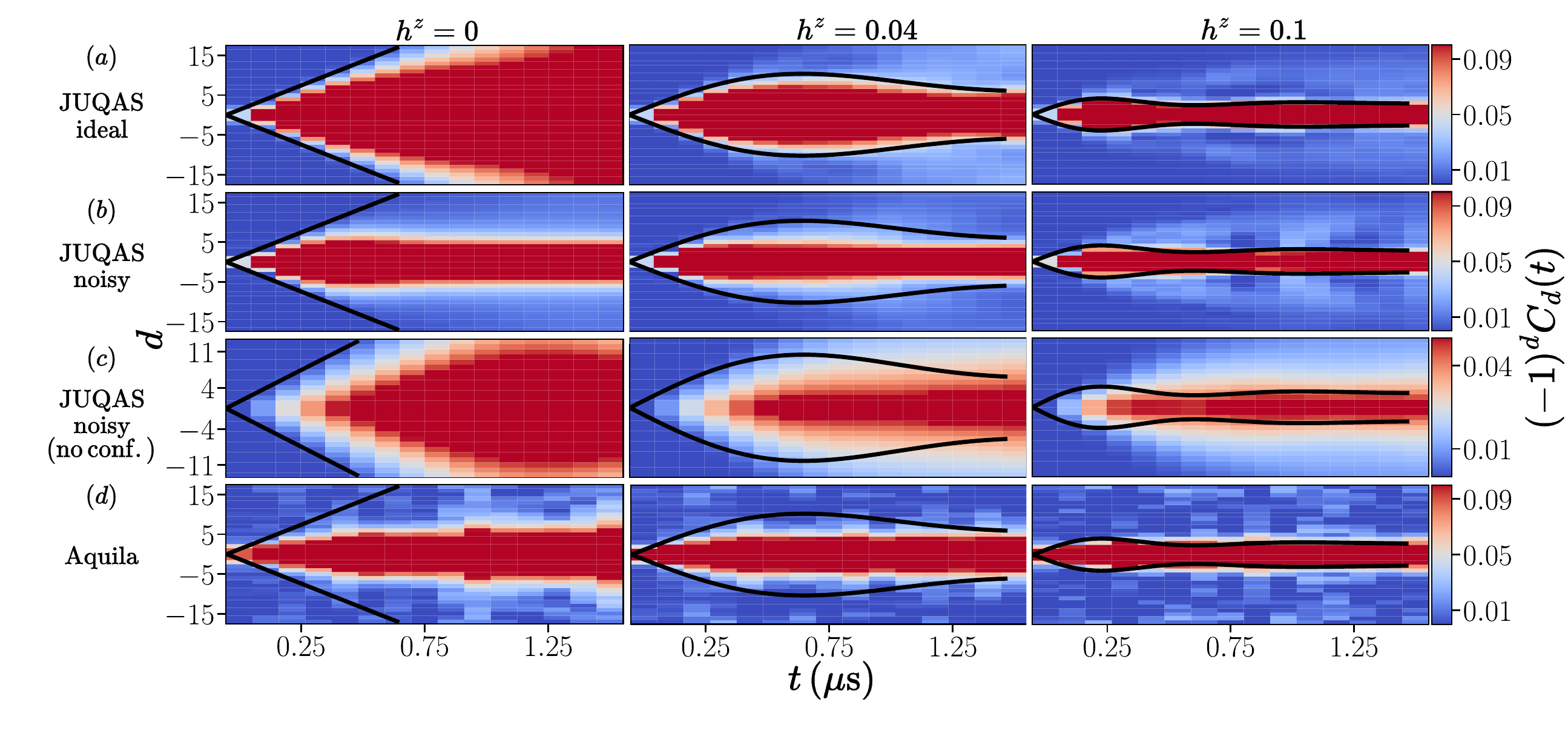}
\caption{
    \textbf{Confinement and noise–induced suppression of correlation spreading.}
    Time evolution of the staggered connected correlations $(-1)^d C_d(t)$
    for a 35-atom chain at three longitudinal-field strengths 
    $h^z = 0$ (left), $0.04$ (middle), and $0.1$ (right). Black curves denote the semiclassical single-meson
    prediction for the maximal correlation front of the ideal confined system. 
    Rows correspond to:
    (a)~ideal coherent JUQAS simulations without noise, showing ballistic spreading at $h^z = 0$
    and progressively stronger confinement-induced suppression as $h^z$ increases.
    (b)~Noisy JUQAS simulations with the experimentally implemented, spatially varying $h_i$ pattern ($...0101010...$) with added noise, that translates to a homogeneous longitudinal field with added noise in the ferromagnetic quantum Ising model interpretation of the Rydberg Hamiltonian.
    Both confinement ($h_z=0.1$) and noise-induced localization ($h_z=0$) restrict the propagation of correlations.
    (c)~Noisy JUQAS simulations with a uniform longitudinal-field pattern ($h_i = 1$) with added noise. The pattern translates to a staggered longitudinal field in the ferromagnetic quantum Ising interpretation of the Rydberg Hamiltonian, thereby removing confinement so that only localization limits the spread; 
    as $h^z$ increases, the magnitude of the noise also increases in the same way as in (b), and localization becomes stronger, but propagation is still suppressed less than in (b).
    (d)~Experimental data from the Aquila neutral-atom quantum simulator, 
    showing qualitatively similar saturation of the correlation front due to noise-induced localization at $h_z=0$ and 
    confinement dominated suppression at $h_z=0.1$.
}
\label{fig:MainResult}
\end{figure*}

Following initialization (see Methods for details) in the Néel state in the original Rydberg Hamiltonian, which becomes the homogeneous polarized state in the ferromagnetic Ising model interpretation, we perform a sudden quench of $(h^x,h^z)$ and monitor the evolution of the connected spin–spin correlations
\[
C_d(t) = \frac1 {D(d)}\sum_i^{D(d)} \langle\sigma_i^z(t)\sigma_{i+d}^z(t)\rangle
- \langle\sigma_i^z(t)\rangle\langle\sigma_{i+d}^z(t)\rangle,
\]
where $D(d)$ is the number of $(i,j)$ pairs in the chain which satisfies $|j-i|=d$.
Fig.~\ref{fig:MainResult} summarizes the experimental and simulated results.
Starting from the fermionic model of fermions in a $1$D chain \cite{scopa2022entanglement}, the expected behaviour after a quench to a small longitudinal field is the formation of fermion pairs with momenta $(-k,k)$ and the emergence of a confining force that bends the short-time ballistic trajectories $d(t,k)\backsimeq 2v(k)t$, where $v(k) = \partial_k \varepsilon(k)$ and $\varepsilon (k) = 2J\sqrt{(\cos k - h^x)^2 + \sin^2 k}$. Focusing on a single pair created at momenta $(-k, k)$ and overlapping initial position, the classical distance is given by \cite{scopa2022entanglement}
\begin{equation}
d(t,k) = \frac{[\varepsilon(k)-\varepsilon(k-2|h^z|\bar\sigma t\mod(k/(|h^z|\bar\sigma))] }{|h^z|\bar\sigma},
\end{equation}
where $\bar\sigma = (1-(h^x)^2)^{1/8}$, $\mathcal K(k) = \tan(\Delta\theta_k/2) + h^z \bar\sigma v(k)/\varepsilon^2(k)$, and $n = \sum_k \frac{|\mathcal K(k)|^2}{1+|\mathcal K(k)|^2}$.
In Fig.~\ref{fig:MainResult}, we show the average distance $\langle d(t,k) \rangle$ as black lines, for three different values of $h^z$, evaluated numerically as
\begin{equation}
\displaystyle\langle d(t,k) \rangle = \dfrac{\int_0^{\pi} d(t,k) n(k)\, dk}{\int_0^{\pi} n(k)\, dk},
\end{equation}
by considering a continuous spectrum for the momentum modes.

Panel~(a) shows coherent JUQAS simulations of the ideal Hamiltonian without noise.  
For $h^z = 0$, correlations spread ballistically, forming a clear light cone associated with freely propagating domain walls.  
As $h^z$ increases, confinement emerges: the longitudinal field generates a linear potential that binds domain walls into meson-like excitations, suppressing their motion.  
At $h^z = 0.1$, the correlation front becomes nearly static, consistent with strong confinement.

Panel~(b) displays noisy JUQAS simulations using the experimentally implemented, spatially varying longitudinal-field pattern 
$(\dots 010101 \dots)$ together with added noise. This translates to a homogeneous longitudinal field with added noise in the ferromagnetic quantum Ising model interpretation of the Rydberg Hamiltonian and initial state is a fully polarized state.
In this configuration, correlations are affected by two mechanisms:  
(i)~noise-induced localization, which dominates at $h^z = 0$ and restricts the propagation to short distances, and  
(ii)~confinement, which becomes increasingly important as $h^z$ grows and strongly suppresses spreading.  
The combined effect of confinement and localization leads to a noticeably stronger reduction of the correlation front than in the ideal case in the intermediate case ($h_z=0.04$) and confinement dominated behavior at the strongest longitudinal field ($h_z=0.1$).

Panel~(c) isolates the effect of localization by removing confinement entirely: we set the longitudinal-field pattern to a uniform value $h_i = 1$ with added noise, so the linear potential binding domain walls is absent. This translates to a staggered longitudinal field with added noise in the ferromagnetic quantum Ising model interpretation of the Rydberg Hamiltonian, while the initial state is still the fully polarized state.
Only noise-induced localization—arising from small spatial fluctuations in $h_i$—limits propagation.  
As $h^z$ increases, this localization becomes stronger and reduces the correlation range, but always less than in panel~(b).  
This demonstrates that confinement is the dominant mechanism halting correlation spreading at sufficiently large $h^z$.

Panel~(d) shows experimental measurements obtained on the Aquila neutral-atom quantum simulator.  
At $h^z = 0$, the correlation front saturates at short distances due to noise-induced localization, qualitatively matching panel~(b).  
At $h^z = 0.1$, confinement dominates the dynamics, leading to a nearly static correlation profile consistent with the confined regime.

Overall, comparing panel~(b)–(d) demonstrates that while noise-induced localization alone restricts correlation spreading to a finite distance, confinement provides an additional and much stronger suppression mechanism, eventually freezing the propagation altogether when $h^z$ is large.  
The black curves represent the semiclassical single-meson prediction for the maximal correlation front in the ideal confined system, accurately capturing the early-time dynamics.

\subsection{Quantitative comparison between emulation and experiment}

\begin{figure}[t]
    \includegraphics[width=1\columnwidth]{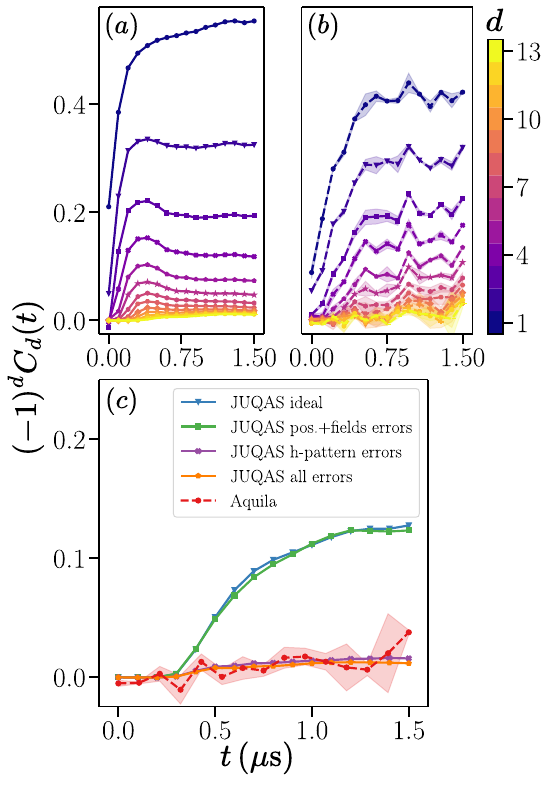}
    \centering
    \caption{
        \textbf{Qualitative comparison between JUQAS emulations and Aquila experiment}. Time evolution of the distance-resolved connected correlations $(-1)^d C_d(t)$ for (a) noisy JUQAS emulations and (b) experimental data from the Aquila Rydberg simulator. (c) Comparison at $d=12$ between ideal emulations, emulations including only atom-position and field errors, emulations including only $h$-pattern errors, emulations including all error sources listed in Table~\ref{table:error specs}, and the experimental results. The noisy emulations employ error amplitudes scaled by 1.5 above the nominal device specifications, yielding the best quantitative agreement with experiment. Shaded regions indicate standard deviations from multiple noise realizations and experimental repetitions. The close agreement across all distances demonstrates that the enhanced noise model captures the effective inhomogeneity and initialization errors of the device, with the $h$-pattern emerging as the dominant error source.
    }
    \label{fig:Juqas_vs_Aquila}
\end{figure}

To quantitatively assess the influence of noise, we compare the distance-resolved correlations $(-1)^dC_d(t)$ between the noisy JUQAS emulations and experimental data, shown in Fig.~\ref{fig:Juqas_vs_Aquila}.  

The results in Fig.~\ref{fig:Juqas_vs_Aquila}(a) show the measurements from Aquila, while Fig.~\ref{fig:Juqas_vs_Aquila}(b) presents the JUQAS emulations, which incorporate all known sources of experimental inhomogeneity. In these simulations, the noise amplitudes are taken to be larger than the nominal device specifications by a factor of 1.5 in order to account for correlated drifts and unmodeled fluctuations. For a fixed distance, Fig.~\ref{fig:Juqas_vs_Aquila}(c) compares emulations with selected noise sources to the real-device data.

The good agreement between the full-error emulations and the experimental results, particularly at larger distances, indicates that the observed saturation of correlations and the reduced amplitude are primarily caused by spatial noise, rather than by intrinsic decoherence or thermalization mechanisms. Among the various noise sources considered, inhomogeneity in the local longitudinal-field profile was identified as the dominant contribution, effectively producing emergent disorder that spatially pins the excitations.

\subsection{System-size dependence}

\begin{figure}[t]
    \includegraphics[width=1\columnwidth]{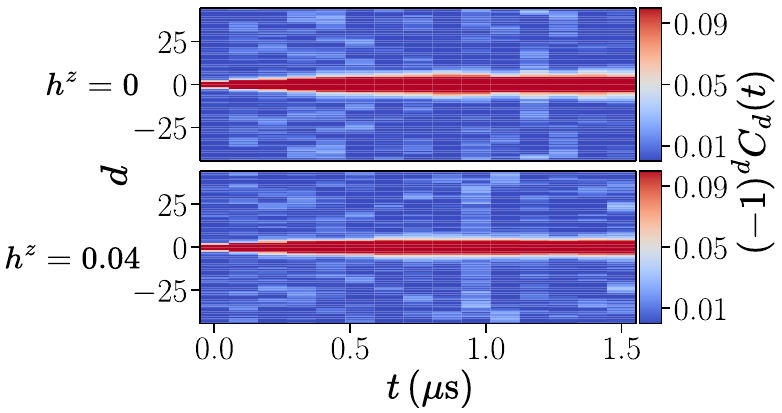}
    \centering
    \caption{
        \textbf{Correlation dynamics in a larger Rydberg chain.}
        Time evolution of the staggered connected correlations $(-1)^d C_d(t)$
        measured on the Aquila device for a chain of $L = 89$ atoms.  
        (up)~Deconfined regime with $h^z = 0$ and (down)~confined regime with $h^z = 0.04$.  
        In both cases, correlations extend only over short distances relative to the system size and quickly saturate, indicating strong decoherence and signal loss consistent with the effects of device noise.  
        The absence of sustained propagation or oscillatory behavior, together with the rapid damping of correlations, suggests that increasing system size leads to a reduced signal-to-noise ratio and enhanced effective localization of excitations.
    }
    \label{fig:large_Rydberg_chain}
\end{figure}

We further explore scaling by performing the same protocol on a longer chain of $L=89$ atoms (Fig.~\ref{fig:large_Rydberg_chain}).  
In both the deconfined ($h^z=0$) and weakly confined ($h^z=0.04$) regimes, correlations extend only over short distances relative to the system size and rapidly saturate.  
The reduced signal-to-noise ratio at larger $L$ amplifies the effect of spatial noise, leading to stronger apparent localization of correlations.  
No oscillatory confinement dynamics are observed, indicating that noise dominates the dynamics on these timescales.  
This scaling behavior confirms that the experimentally accessible regime is one where confinement and spatial noise act cooperatively to suppress correlation spreading and preserve memory of the initial state.

\subsection{Interplay of confinement and localization}

The combination of quantum simulation and coherent emulation establishes a unified picture:  
confinement introduces an energetic cost for domain-wall separation, while spatial and temporal noise generate an effective localization length that limits correlation propagation.  
In the absence of noise, domain walls form coherent mesonic oscillations; in its presence, the same excitations become pinned and fail to propagate beyond a finite distance.  
Together, these mechanisms yield a continuous crossover from noise-induced localization to coherent confinement with increasing the strength of confinement from zero ($h_z=0$) to strong confinement ($h_z=0.1$)—two complementary routes to non-ergodic dynamics in Rydberg spin chains.

\section{Discussion and Outlook}

The results presented here establish a detailed understanding of how confinement and noise jointly shape the non-equilibrium dynamics of Rydberg spin chains.  
In the absence of imperfections, the longitudinal-field term of the transverse-field Ising model generates confinement, binding domain-wall excitations into mesons whose motion produces the coherent light-cone patterns observed in ideal simulations.  
On real quantum hardware, however, additional spatial and temporal noise—such as inhomogeneous detunings, atomic-position fluctuations, and laser-amplitude variations—introduces effective disorder that limits the motion of these bound states.  
Consequently, the experimental dynamics exhibit an initial spread followed by a rapid saturation of correlations, in quantitative agreement with noisy JUQAS emulations that incorporate the same imperfections.

A key insight emerging from our results is that, although near-term devices inevitably exhibit noise-induced localization, confinement remains a tunable and robust mechanism that can dominate the dynamics when the longitudinal field is sufficiently strong.  
Localization alone restricts propagation to a finite distance, but the addition of confinement suppresses motion even further, ultimately freezing the correlation front at large $h^z$.  
This demonstrates that the underlying confinement physics is not washed out by hardware noise; rather, by tuning $h^z$ one can access a regime where confinement overwhelms localization and dictates the system's behavior.

More broadly, our results bridge concepts typically associated with distinct physical regimes: interaction-induced confinement in clean systems and localization in disordered or noisy settings.  
The coexistence of these effects on programmable quantum hardware provides a new means for studying constrained many-body dynamics beyond idealized theoretical limits.  
By varying field strengths and device noise levels, one can move continuously from localization-limited spreading to confinement-dominated frozen dynamics, enabling systematic exploration of how these two mechanisms interact.

Looking ahead, several natural extensions arise.  
First, experiments employing controlled spatial detuning gradients or engineered disorder would allow quantitative tests of how confinement length and localization length compete.  
Second, noise-mitigation or error-suppression strategies tailored to analog quantum simulators could help disentangle intrinsic confinement physics from hardware-induced localization and extend the noiseless evolution window.  
Finally, extending these studies to two-dimensional Rydberg arrays or ladder geometries would enable exploration of confinement and localization in higher dimensions, where string tension, domain-wall topology, and emergent gauge constraints are expected to have a central impact.

In summary, the combined numerical and experimental results show that Rydberg quantum simulators naturally realize a regime where confinement and noise-induced localization coexist, but where confinement can be tuned to become the dominant mechanism suppressing transport.  
This regime provides a powerful platform for probing mechanisms of ergodicity breaking and for developing quantum technologies that harness, rather than merely tolerate, the complex dynamics emerging at the interface between coherence and noise.

\section{Methods}

\subsection{Atomic Register}

In configuring the atomic register for the experiment, it was necessary to introduce slight turns in the atomic chain, as the physical constraints of the quantum device prevented an indefinitely extended linear arrangement.  
The main challenge associated with these turns stems from the next–next–nearest-neighbor interactions experienced by atoms at the bends, which are stronger than those in the linear sections of the chain.  
If these interactions become too large, they contribute non-negligibly to the system’s dynamics, thereby breaking the regime in which the interaction strengths can be treated as approximately homogeneous along the chain.

To quantify this effect, we performed numerical emulations for turn angles ranging from $0^\circ$ to $90^\circ$.  
Our simulations indicate that for a turn angle of $60^\circ$ and an interatomic spacing of $a = 6.0\,\mathrm{\mu m}$, the impact of next–next–nearest-neighbor interactions remains negligible.  
Under these conditions, the interaction strengths are effectively homogeneous throughout the register.

Finally, we note that the chosen register employs an open-chain configuration.  
Due to the absence of periodic boundary conditions, the two atoms at the edges experience only half of the interaction contributions compared to those in the bulk.  
This asymmetry affects both the $\sigma_i^z$ and $\sigma_i^z\sigma_{i+1}^z$ terms in the effective Hamiltonian.

\subsection{Initial State Preparation}

The first stage of the quantum evolution schedule is dedicated to preparing the antiferromagnetic Néel state $| 10\ldots01\rangle$, where only the atoms at odd lattice positions occupy the Rydberg state. One way to achieve this configuration, would be to apply a transverse field selectively to the atoms in odd positions, inducing an effective rotation of their quantum state on the Bloch sphere from the initial ground state $|0\rangle$ at $t = 0$ to the excited state $|1\rangle$. This transformation can be implemented using the rotation operator $R_x(\theta) = e^{-i\frac{\theta}{2} \sigma^x}$ with $\theta = \pi$.
If a local transverse field were available, this rotation could be realized by applying the evolution operator $U(t_{\pi}) = e^{-i \int_0^{t_\pi}\sum_{i_{\mathrm{odd}}}\Omega_\pi t \sigma^x_i dt}$, where we chose the maximum allowed Rabi frequency $\Omega_\pi=\Omega_\mathrm{max} = 2.5 \times 2\pi \,\mathrm{MHz}$ to minimize the preparation time. This choice yields $t_\pi = 0.2 \,\mathrm{\mu s}$.
However, due to the lack of local transverse field control in the experimental setup, the Néel state must instead be prepared using the local detuning term in the Hamiltonian \eqref{eq:HRydNN}, where the detuning pattern is defined as $h_i = 0$ for odd sites and $h_i = 1$ for even sites. During the application of the global Rabi frequency $\Omega(t)$, we simultaneously apply the lowest available local detuning, $\Delta_\mathrm{loc, min} = -7.5 \times 2\pi \,\mathrm{MHz}$, to suppress excitations at even sites.
A final consideration regarding the initialization schedule is that, in the actual device, both $\Omega(t)$ and $\Delta_\mathrm{loc}(t)$ must begin from zero. To reach the target values, we implement a quench of duration $t_\mathrm{quench} = 0.05 \,\mathrm{\mu s}$, the minimum time step allowed by the device, to ramp up $\Omega_\pi$. Due to hardware constraints, a longer duration of $t_\mathrm{loc}= 0.2\,\mathrm{\mu s}$ is required to reach $\Delta_\mathrm{loc, min}$, as tests on the actual device showed that shorter times degrade the initial state quality. Moreover, considering that, with the presence of local detuning, the full evolution operator is $U(t_\mathrm{prep}) = \mathcal{T} \Big[ e^{-i \int_0^{t_\mathrm{prep}}\sum_i (\frac{\Omega (t)}{2} \sigma^x_i + h_i\frac{\Delta_\mathrm{loc}(t)}{2}\sigma_i^z)t dt} \Big]$, where $\mathcal{T}$ denotes the time-ordering operator and $t_\mathrm{prep} = 2t_\mathrm{loc} +2t_\mathrm{quench}+ t_\pi$, we performed real device tests for different values of $t_\pi$ and $\Omega_\pi$ with values in a range around $t_\pi = 0.2 \,\mathrm{\mu s}$ and $2\times 2\pi \,\mathrm{MHz}$ respectively. These emulations indicated that a reduced duration of $t_\pi = 0.15 \,\mathrm{\mu s}$ and $\Omega_\pi = 2.05 \times 2\pi \,\mathrm{MHz}$ yields a higher-fidelity Néel state compared to the initially chosen ones. The protocol for the initial state preparation is represented in Fig. \ref{Fig:protocol}(b).

\subsection{Quantum Simulation Pulse Sequence}

Following the initial state preparation, we implemented the second phase of the field schedule for the quantum simulation, aimed at observing confinement dynamics in the real-time evolution of the connected correlations $C_d(t)$. As outlined in \cite{kormos2017real}, within the framework of the ferromagnetic Ising Hamiltonian in Eq. \eqref{eq:HIsiFM1}, the protocol consists of a quench of the transverse and longitudinal fields to selected values $h^x$ and $h^z$, followed by an evolution under these fixed parameters for an arbitrary duration. The mapping between the ferromagnetic Ising Hamiltonian in Eq. \eqref{eq:HIsiFM1} and the Rydberg Hamiltonian in Eq. \eqref{eq:HRyd} was previously established by rewriting the latter in the form of Eq. \eqref{eq:HRydNN}.

For given values of $h^x$ and $h^z$, the corresponding parameters in the Rydberg system, namely the global Rabi frequency $\Omega$, the global detuning $\Delta_\mathrm{glo}$, and the local detuning $\Delta_\mathrm{loc}$, are obtained through the relations $\frac{2\Omega}{U_{i,i+1}} = h^x$ and $\bigg(-\dfrac{2\Delta_\mathrm{glo} }{U_{\mathrm{i,i+1}}}
-\dfrac{2h_i\Delta_\mathrm{loc} }{U_{\mathrm{i,i+1}}} +2\bigg) = -h^z_i(-1)^i$. To obtain the corresponding staggered values of $\pm h^z_i$ for odd and even sites in the $\sigma_i^z$ terms for the Rydberg Hamiltonian, we have used the $h_i$ values in \eqref{eq:HRydNN}  to be $h_i = 1$ if $i$ is even and $h_i = 0$ if $i$ is odd. With this we obtain that, for odd sites $-\dfrac{2\Delta_\mathrm{glo} }{U_{\mathrm{i,i+1}}}+2 = -h^z_i$ and for even sites $-\dfrac{2\Delta_\mathrm{glo} }{U_{\mathrm{i,i+1}}}-\dfrac{2\Delta_\mathrm{loc} }{U_{\mathrm{i,i+1}}} +2 = h^z_i$ and so this will lead to $\Delta_\mathrm{glo} = U_{i,i+1}(1+h^z_i/2)$ and $\Delta_\mathrm{loc} = -U_{i,i+1}h^z_i$.

In this study, we set $h^x = 0.25$, corresponding to a Rabi frequency $\Omega = 2.31 \times 2\pi \,\mathrm{MHz}$. For the regime where no confinement effects are expected, the longitudinal field magnitude $h^z$ must be set to zero, yielding the parameter values $\Delta_\mathrm{glo} =  18.5\times 2\pi \,\mathrm{MHz}$ and $\Delta_\mathrm{loc} = 0$. Conversely, to probe the confinement regime, we choose a nonzero longitudinal field, setting $h^z = 0.04$, which corresponds to $\Delta_\mathrm{glo} = 18.12 \times 2\pi \,\mathrm{MHz}$ and $\Delta_\mathrm{loc} = -0.74 \times 2\pi \,\mathrm{MHz}$.
The fields are quenched to respective values in the shortes time available in the device $t_\mathrm{quench} = 0.05\,\mathrm{\mu s}$, then kept constant for $t_\mathrm{sim} = 1.5\,\mathrm{\mu s}$ and then quenched back to zero values at the end of the evolution. The protocol for the quantum simulation is represented in Fig. \ref{Fig:protocol}(b).

\subsection{Noisy coherent emulation}

To gain insights into the behavior of the real device, we performed coherent numerical emulations using the Jülich Quantum Annealing Simulator (JUQAS). The time evolution of the system was emulated by solving the time-dependent Schrödinger equation (with $\hbar =1$)

\begin{equation}
i\frac{\partial}{\partial t}|\psi(t)\rangle =H(t)|\psi(t)\rangle
\end{equation}

numerically, under the assumption of an ideal closed system. Here, $|\psi(t)\rangle$ represents the state vector, while the Hamiltonian governing the system dynamics is given by

\begin{equation}
\begin{split}
{H}(t) = \sum\limits_{i} \Big[\big( f_i^x(t)h_i^x\sigma_i^x + f_i^y(t)h_i^y\sigma_i^y + f_i^z(t)h_i^z\sigma_i^z \big) + \\
+\sum\limits_{j<i} \big( F_{ij}^x(t)J_{ij}^x \sigma_i^x\sigma_j^x + F_{ij}^y(t)J_{ij}^y \sigma_i^y\sigma_j^y + F_{ij}^z(t)J_{ij}^z \sigma_i^z\sigma_j^z \big)\Big].    
\end{split}
\end{equation}

This general form of the Hamiltonian enables the emulation of both quantum annealers and quantum simulators. The numerical approach employed for solving the Schrödinger equation is the Suzuki-Trotter product-formula algorithm \cite{suzuki1976,suzuki84,trotter59,deraedt87,huyghebaert1990}, which allows for full state-vector emulation. The implementation used in our emulator supports decomposition either into single- and two-qubit terms or into $\sigma^x$, $\sigma^y$, and $\sigma^z$ terms.

The emulator is optimized for GPU acceleration using OpenACC, and due to the exponential memory requirements for storing the state vector, distributed memory over multiple GPUs is required for large system sizes ($\gtrapprox 30$ qubits, depending on GPU memory capacity). For inter-GPU communication, we employ CUDA-aware MPI (Message Passing Interface), utilizing a communication scheme similar to that implemented in the Jülich Universal Quantum Computer Simulator (JUQCS) \cite{deraedt07,deraedt18,Willsch2022_gpu}.

The results of ideal emulations for a chain of $35$ atoms, shown in Fig.~\ref{fig:MainResult}(a), reveal a significant discrepancy when compared to experimental results obtained on the real device (see Fig.~\ref{fig:MainResult}(d)). This deviation highlights the presence of hardware-induced imperfections in the actual quantum system.

To achieve a more realistic emulation that closely reproduces the behavior of the physical device, we incorporated known hardware imperfections based on the device specifications. The considered sources of errors are showed in Tab. \ref{table:error specs}.

\begin{table}
\begin{center}
\begin{tabular}{ |c|c| } 
\hline
\textbf{Quantity} & \textbf{Error}\\
\hline
  $\;$Atomic coordinates & Absolute error of $0.1 \,\mathrm{\mu m}$ $\;$\\
 \hline
 $\Omega$ & Relative error of $2\%$ \\
 \hline
 $\Delta_\mathrm{glo}$ &  $\;$Absolute error of $1\, \mathrm{MHz}$$\;$\\
\hline
$\Delta_\mathrm{loc}$ & Relative error of $2\%$\\
\hline
$h$-pattern & Error of $0.1$\\
\hline
\end{tabular}
\caption{\textbf{Error specifications for Aquila}.
“Atomic coordinates’’ denote the precision limits in positioning individual atoms during register preparation.
The parameters $\Omega$ and $\Delta_\mathrm{loc}$ are affected by relative deviations from their programmed values, whereas $\Delta_\mathrm{glo}$ exhibits absolute offsets as well as site-dependent inhomogeneity.
 Finally, the implemented $h$-pattern also carries an absolute per-site error.}
\label{table:error specs}
\end{center}
\end{table}



Simulations incorporating the device-reported error values did not reproduce the experimental behavior. To assess the impact of imperfections, we systematically increased the error amplitudes beyond the specified values. We found that scaling the reported errors by a factor of approximately $1.5$ yields emulation results that closely match the experimental data, as shown in Fig.~\ref{fig:MainResult}(b). This enhanced error requirement is likely linked to the use of local detuning, which is expected to amplify the contribution of underlying imperfections.

A detailed comparison between experimental data and noisy JUQAS emulations for distances up to $d=13$ is presented in Fig.~\ref{fig:Juqas_vs_Aquila}. Figure~\ref{fig:Juqas_vs_Aquila}(a) shows noisy emulations including all error sources, while Fig.~\ref{fig:Juqas_vs_Aquila}(b) displays the corresponding experimental results, with error bars representing standard deviations over three repetitions of the experiment.

To isolate the dominant noise mechanisms, we performed additional emulations including only selected subsets of the error sources listed in Table~\ref{table:error specs}. Figure~\ref{fig:Juqas_vs_Aquila}(c) compares JUQAS results with Aquila data for the $h^z=0$ case at $d=12$, using the same scaling factor of $\backsim 1.5$. We show (i) ideal emulations, (ii) emulations with only atom-position and field errors, (iii) emulations with only $h$-pattern errors, and (iv) emulations including all error sources. The comparison indicates that the $h$-pattern inhomogeneity is the primary contributor to deviations from ideal behavior.



\section{Data Availability}

Quantum simulation data obtained from QuEra's Aquila quantum simulator were deposited to Zenodo and are available at the following URL: \url{https://doi.org/10.5281/zenodo.17739603}.\\


\section{Acknowledgments}

J.V., A.B.R., J.A.M-B. acknowledge support from the project Jülich UNified Infrastructure for Quantum computing (JUNIQ) that has received funding from the German Federal Ministry of Education and Research (BMBF) and the Ministry of Culture and Science of the State of North Rhine-Westphalia. A.B.R. acknowledges support from the project HPCQS (101018180) of the European High-Performance Computing Joint Undertaking (EuroHPC JU).

We acknowledge support from Amazon Web Services (AWS) through the provision of Amazon Braket credits used for quantum computing experiments on QuEra Aquila.

We are also grateful for the discussions with Gianluca Lagnese.


\section{Funding}

This study was funded by the Slovenian Research Agency grants P1-0040 and P1-0416, ERC AdG grant ‘HIMMS’, Jülich UNified Infrastructure for Quantum computing (JUNIQ), German Federal Ministry of Education and Research (BMBF), the Ministry of Culture and Science of the State of North Rhine-Westphalia, and the project HPCQS (101018180) of the European High-Performance Computing Joint Undertaking (EuroHPC JU). The funder played no role in study design, data collection, analysis and interpretation of data, or the writing of this manuscript. 


\section{Author Contributions}

J.V. developed the idea and supervised the work, A.B.R. and J.A.M-B. performed and analyzed the quantum simulations, A.B.R. performed the emulation of quantum simulations, and all authors wrote the manuscript.


\section{Competing interests}

The authors declare no competing interests.


\bibliographystyle{apsrev4-2}
\bibliography{biblio.bib}

\end{document}